# Adjustable current-induced magnetization switching utilizing interlayer exchange coupling

*Yu Sheng, Kevin William Edmonds, Xingqiao Ma, Houzhi Zheng and Kaiyou Wang\**

Prof. K. Wang, Prof. H. Zheng

State Key Laboratory of Superlattices and Microstructures

Institute of Semiconductors

Chinese Academy of Sciences

Beijing 100083, P. R. China

Email: kywang@semi.ac.cn

Y. Sheng, Prof. X. Ma

Department of Physics

University of Science and Technology Beijing

Beijing 100083, P. R. China

Dr. K. W. Edmonds

School of Physics and Astronomy

University of Nottingham

Nottingham NG7 2RD, United Kingdom.

Prof. K. Wang, Prof. H. Zheng

College of Materials Science and Opto-Electronic Technology

University of Chinese Academy of Science

Beijing 100049, P. R. China

Prof. K. Wang

Center for Excellence in Topological Quantum Computation

University of Chinese Academy of Science

Beijing 100049, P. R. China.

Keywords: spintronictcs, spin-orbit torque, field-free switching, interlayer exchange coupling

Electrical current-induced deterministic magnetization switching in a magnetic multilayer structure without external magnetic field is realized by utilizing interlayer exchange coupling. Two ferromagnetic Co layers, with in-plane and out-of-plane anisotropy respectively, are separated by a spacer Ta layer, which plays a dual role of inducing antiferromagnetic interlayer coupling, and contributing to the current-induced effective magnetic field through the spin Hall effect. The current-induced magnetization switching behavior can be tuned by pre-magnetizing the in-plane Co layer. The antiferromagnetic exchange coupling field increases with decreasing thickness of the Ta layer, reaching 630 ±5 Oe for a Ta thickness of 1.5nm. The magnitude of the current-induced perpendicular effective magnetic field from spin-orbit torque is 9.2 Oe/($10^7$Acm$^{-2}$). The large spin Hall angle of Ta, opposite in sign to that of Pt, results in a low critical current density of $9 \times 10^6$A/cm$^2$. This approach is promising for the electrical switching of magnetic memory elements without external magnetic field.

## 1. Introduction

Current-induced magnetization switching and magnetic domain wall (DW) motion in materials with perpendicular magnetic anisotropy (PMA), as a candidate scheme for the next generation of data storage and logic devices, has been widely investigated in recent years.[1–9] Compared with magnetization switching by magnetic field, current-induced switching enables higher storage density, faster writing speed, and lower energy consumption. In normal metal (NM)/ferromagnetic metal (FM)/normal metal stacks, the switching caused by spin-orbit torques (SOTs) commonly consists of a damping-like torque along $\vec{m} \times (\vec{\sigma} \times \vec{m})$ and a field-like torque along $\vec{m} \times \vec{\sigma}$, where $\vec{m}$ is the magnetic moment, and $\vec{\sigma}$ is the spin polarization of spin current.[3,6,10–12] The field-like torque, mainly arising from the Rashba effect caused by the asymmetry normal to the film, can affect the threshold switching current.[13] The damping-like torque together with an in-plane magnetic field are commonly considered as the main factors determining the switching direction.[6,10] The damping-like torque in ferromagnetic layers comes from the absorption of electron spins induced by spin Hall effect in adjacent normal metal layers like Pt, Ta, Hf and W,[3,6,12,14–19] or antiferromagnetic materials (AFM) layers like PtMn and IrMn.[4,20,21]

For deterministic switching of perpendicular magnetization using spin-orbit torques, it is necessary to break the symmetry between up and down magnetization directions, commonly by using an in-plane magnetic field.[10,16,22] As a replacement for the external magnetic field, switching assisted by effective in-plane magnetic fields was demonstrated using a polarized ferroelectric substrate or stacks with interlayer coupling using antiferromagnetic layers.[3,4,21,23] Antiferromagnetic layers can provide an effective in-plane field by exchange bias, and can also act as a spin current source through the spin Hall effect.[7,10,11,22] However, it is difficult to switch the AFM effectively due to its insensitivity to magnetic field, so the interlayer exchange cannot be used tune the current-induced magnetization switching behavior. Current-induced magnetization switching without external magnetic field has also been achieved in stacks of Pt/FM/Ru/FM/AFM using interlayer exchange coupling (IEC), where the AFM layer is used to fix the magnetization orientation of the adjacent FM layer.[23] The Ru layer in these stacks enables strong IEC, but has small spin Hall angle ($\theta_{SH}$) of 0.0056 of the same sign of Pt layer,[24] while $\theta_{SH}$ of Pt is 0.013 – 0.1,[26,27]. Therefore the Ru layer generates negligible spin current when charge current is injected into the device, and only increases the power consumption.

In this work, field-free current-induced magnetization switching is achieved in a Ta/Pt/Co/Ta/Co/Pt structure. The two ferromagnetic Co layers with different thickness are antiferromagnetically coupled through the Ta spacer layer. The thinner bottom Co layer has perpendicular magnetic anisotropy and the thicker top Co layer has in-plane magnetic anisotropy. This removes the need for the AFM layer used in previous reports.[23] The magnetization orientation of the in-plane Co layer was set by external magnetic field along the current direction before measurement. This pre-magnetizing step directly affects the current-induced switching direction, as described below. The spacer layer Ta has a large spin Hall angle of 0.07 of opposite sign to that of Pt, [26] which can effectively assist current-induced magnetization switching.[19] The threshold current density for switching the magnetization is $9 \times 10^6$ A/cm$^2$ for a device with 1.5 nm thick Ta spacer layer, corresponding to a significant reduction in the current density compared to the previous report using a Ru spacer.[23] With the top Co layer pre-magnetized along the current direction, deterministic current-induced switching of the bottom Co layer magnetization can be realized. Reversing the top layer magnetization reverses the direction of current-induced magnetization switching. The coupling field is as large as 630 ±5 Oe when the spacer Ta is 1.5 nm thick, which decreases with increasing the thickness of the spacer layer. Current-induced shifts of the magnetization hysteresis loops indicate a perpendicular effective field of 9.2 Oe/($10^7$Acm$^{-2}$).

## 2. Results and discussion

### 2.1. Properties of ferromagnetic stacks and Hall bar devices

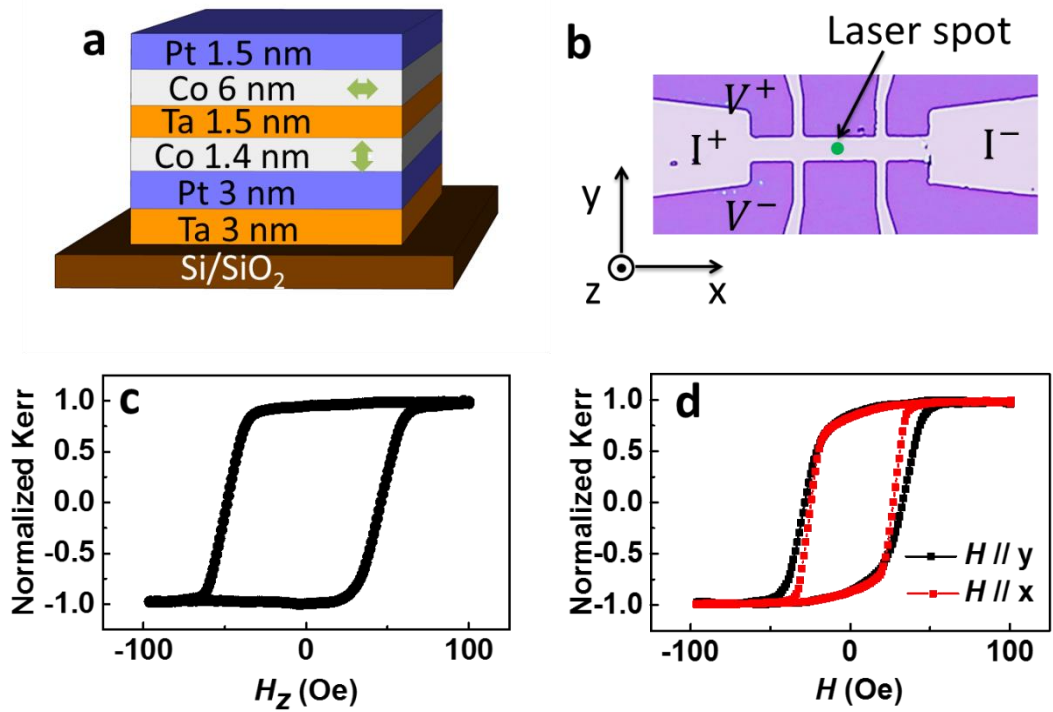

**Figure 1.** Properties of stacks and Hall bar device. (a) Schematic of the stacks used to fabricate into Hall bar to investigate the field-free current-induced magnetization switching. Green double headed arrows indicate the easy axis of the two ferromagnetic

layers. (b) Optical microscope image of a device. Both magnetic-optical Kerr effect and anomalous Hall effect are used to detect the magnetization of device. (c) and (d) magnetization hysteresis loops measured using a MOKE magnetometer in polar and longitudinal configurations, respectively.

Material stacks studied in this work are made up of Ta(3)/Pt(3)/Co(1.4)/Ta($t_{Ta}$)/Co(6)/Pt(2) from bottom to top (thickness in nanometers) with $t_{Ta}$=1.5, 2, 3 and 5 nm, respectively (**Figure 1a**). Sample Ta(3)/Pt(3)/Co(1.4)/Ta(1)/Pt(2) without top ferromagnetic layer is also studied as reference. All layers were sputter-deposited at room temperature on a thermally oxidized Si/SiO$_2$ substrate. The films were patterned into Hall bar devices with conductive channel width of 10 $\mu$m using standard photolithography and lift-off.

Figure 1b shows an optical micrograph of a typical Hall bar along with the definition of the coordinate system. By detecting the magnetic-optical Kerr effect (MOKE) of the device while sweeping the out-of-plane magnetic field, we find that the 1.4nm bottom Co layer exhibits square out-of-plane magnetization loop as shown in Figure 1c with full remnant magnetization, due to the interface magnetic anisotropy arising from the Pt/Co and Co/Ta interfaces [19,28]. With increasing the thickness of the FM layer, the interface magnetic anisotropy becomes relatively weak. The 6nm top Co layer has the magnetic easy axis in the plane with rather large remnant magnetization along both *x* and *y* directions, as shown by the longitudinal MOKE magnetization loop in Figure 1d.

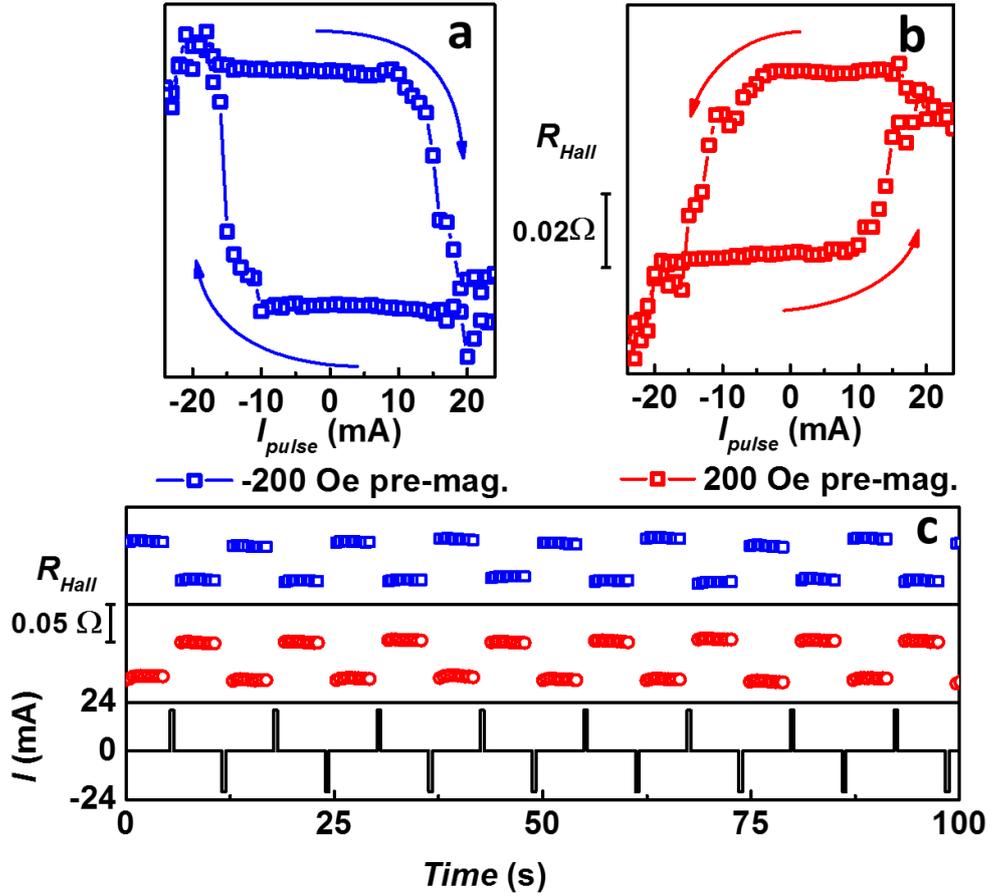

**Figure 2.** Deterministic current-induced magnetization switching under zero magnetic field. Anomalous Hall resistance ($R_{Hall}$) was measured to normal component of magnetization of device with $t_{Ta}$=1.5 nm. (a) $R_{Hall}$ versus $I_{pulse}$ after the device was pre-magnetized by -200 Oe along $x$ axis. (b) $R_{Hall}$ versus $I_{pulse}$ with 200 Oe pre-magnetized. (c) $R_{Hall}$ versus time, measured after device was pre-magnetized by -200 Oe (up panel) and -200 Oe (middle panel), with positive and negative current pulses (bottom panel) applied.

**2.2. Deterministic switching induced by current under zero magnetic field**

**Figure 2** demonstrates the current-induced magnetization switching of the device under zero magnetic field. Firstly, the magnetization state of the top Co layer was set by applying a magnetic field $H_x$ along the *x*-axis of -200 Oe (Figure 2a) or +200 Oe (Figure 2b).

Then the magnetic field was set to zero, and current pulses ($I_{pulse}$) of width 400 ms and varying amplitude were applied to the device. The Hall resistance ($R_{Hall}$) was measured after each pulse at a low current of 0.1 mA to probe the magnetization state of the bottom Co layer. As shown in Figure 2a, a clockwise magnetization switching induced by the electrical current was observed after pre-magnetizing with $H_x = $ -200 Oe, where the current sweeping from positive to negative favors the up magnetization and that from negative to positive favors the down magnetization. Pre-magnetizing with $H_x = $ +200 Oe results in the opposite direction of magnetization switching, as shown in Figure 2b. The critical current density, which was calculated utilizing the current where $R_{Hall}$ is half switched, is $9 \times 10^6 A/cm^2$. Compared with the previously reported device with Ru spacer[23], the threshold current was reduced by around a factor of 3.

Finally, we investigated the magnetization switching under repeated positive and negative 20mA current pulses. As shown in Figure 2c, reproducible switching from +*z* to −*z* (-*z* to +*z*) direction is induced using a negative (positive) current pulse after pre-magnetizing with $H_x$ = -200 Oe, and vice versa for $H_x$ = +200 Oe. Very good repeatability of the current-induced switching was observed over more than 1000 cycles, indicating the high reliability of this type of device structure. However, no deterministic current-induced magnetization switching was observed if the system was pre-magnetized along *y* direction. For the reference sample, without the top Co layer, current-induced magnetization switching was only observed in the presence of an external magnetic field $|H_x|\geq 600$ Oe (See Supporting information Figure S2).

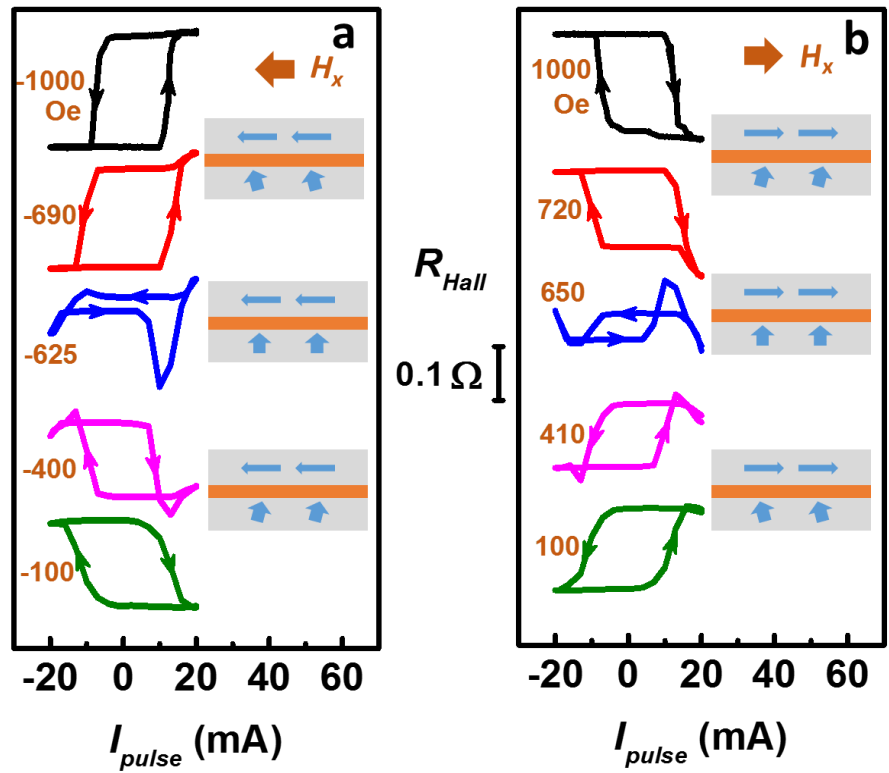

**Figure 3.** Current-induced magnetization switching under different in plane magnetic field. Anomalous Hall resistance of device with $t_{Ta}=1.5$ nm as a function of the injected current pulse measured at different external fields $H_x$, with magnetic field of -1000 Oe, -690 Oe, -625 Oe, -400 Oe and -100 Oe for (a) and with magnetic field of 1000 Oe, 720 Oe, 650 Oe, 410 Oe, and 100 Oe for (b). Schematics inserted illustrate the magnetization configurations of two FM layers under different magnetic field $H_x$, in which the magnetization of bottom Co layer is set to be "up", and is tilted under external field $H_x$ and coupled field.

**2.3. Determination of the interlayer exchange coupling field.**

To understand how the top magnetic Co layer affects the current-induced magnetization switching, we investigate the current pulse-induced switching of the anomalous Hall resistance under different fixed $H_x$. As shown in **Figure 3a**, with fixed magnetic fields of -1000 Oe and -690 Oe applied in the current direction, an anti-clockwise current-induced magnetization switching was observed, which is similar to that of the reference sample (See the Supporting Information Figure S2). With -625 Oe applied, the current-induced magnetization loop becomes much smaller, indicating very weak deterministic switching. Interestingly, with further decreasing the external magnetic field, the deterministic current-induced magnetization switching was observed again. However, opposite sign (clockwise) of current-induced magnetization switching was observed for external magnetic fields of -400 Oe and -100 Oe, compared to that of the 1000 Oe case. The sign change of the current-induced magnetization switching cannot be explained by the stray field from the top Co layer, which is antiparallel to the top layer magnetization with amplitude smaller than 0.001 mT, nor by the stray field created by correlated surface roughness because the Néel orange-peel mechanism always favors parallel alignment of the layers[23].

We ascribe the sign change of the current-induced magnetization switching to the tilt of the bottom layer magnetization resulting from the interplay of the external magnetic field and the antiferromagnetic interlayer exchange coupling. When the external magnetic field is larger than the exchange coupling field, the orientation of the current-induced magnetization switching is fully determined by the external magnetic field direction. When the external magnetic field is comparable to the exchange coupling field, the current-induced magnetization switching has no preferred direction due to the weak total effective magnetic field. When the external magnetic field is much smaller than the exchange coupling field, the current-induced magnetization switching is determined by the exchange coupling field. Hence, the current-induced magnetization switching changes sign with the direction change of the total effective magnetic field in the current orientation. With opposite external magnetic fields applied (Figure 3b), the reversed current-induced magnetization switching loops were observed, which further demonstrates the important role played by the antiferromagnetic exchange coupling between the two ferromagnetic layers. The antiferromagnetic exchange coupling field was determined to be 630 ±5 Oe (details see the Supporting information Figure S3) for $t_{ta}$ = 1.5nm, decreasing with increasing spacer layer thickness.

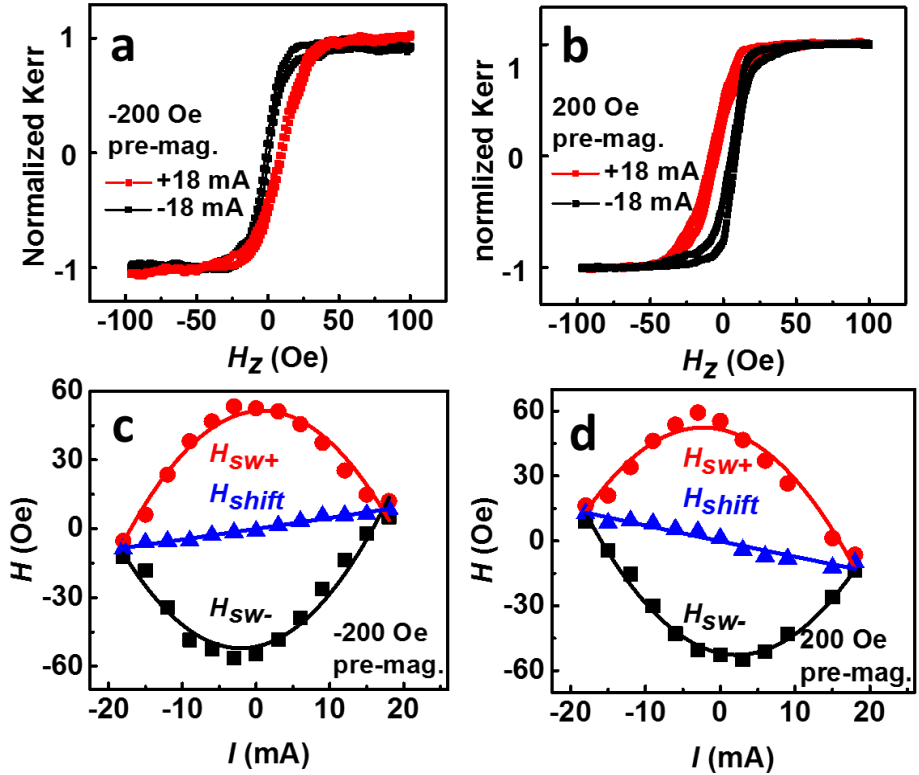

**Figure 4.** Perpendicular effective field induced by current. a), b) Perpendicular magnetization measured by polar MOKE against out-of-plane magnetic field, while currents of +/- 18mA are applied to the devices with zero in-plane external magnetic field. Red (black) circles indicate positive (negative) current. c), d) $H_{sw-}$, $H_{sw+}$ and $H_{shift}$ change with current, the full lines are the fit curve using function $H_{sw-(+)}=H_{0sw-(+)}+(-)\alpha\rho J^2+\beta J$. Before the polar MOKE measurements, the device was magnetized by -200 Oe [a), c)] or +200 Oe [b), d)].

## 2.4. Perpendicular effective field induced by current.

To obtain the relation between the current-induced effective field and the electrical current, we measured the perpendicular magnetization hysteresis loops by polar MOKE at different fixed positive/negative electrical currents. After pre-magnetizing with -200

Oe in-plane external magnetic field, as shown in **Figure 4a**, the magnetization loops shift to the negative (positive) fields with current of -18mA (+18mA) applied. With the sample pre-magnetized by $H_x = +200$ Oe, opposite behavior was observed as shown in Figure 4b. We extracted the switching fields with magnetization from -z to +z ($H_{sw+}$) and +z to -z ($H_{sw-}$), which are shown in Figure 4c and d. The magnitudes of both $H_{sw-}$ and $H_{sw-}$ decrease with increasing the current intensity due to Joule heating. It is worth noting, as shown in Figure 4c, the extreme value for $H_{sw-}$ is at $I<0$ and $H_{sw+}$ is at $I>0$ for -200 Oe pre-magnetized situation. However, opposite behavior was observed for +200 Oe pre-magnetized situation shown in Figure 4d. As a result, the switching field contributed from both SOTs and Joule heating can be expressed as $H_{sw-(+)}=H_{0sw-(+)}+(-)\alpha\rho J^2+\beta J$, where the first term in the right side $H_{0sw-(+)}$ is the switching field from +z to -z (-z to +z) at room temperature without electrical current applied, $J$ is the averaged current density, $\rho$ is the averaged resistivity of the device, the second term $(-)\alpha\rho J^2$ is the dependence of $H_{sw}$ on the Joule heating with the parameter $\alpha$, and the third term is the strength of the current-induced perpendicular effective magnetic field from SOTs with parameter $\beta$. The Joule heating is an even function of the current, reducing the barrier for the magnetization flipping from one state to another and causing both the $H_{sw+}$ and $H_{sw-}$ switching fields to decrease. However, the effective field induced by SOTs depends on the directions of both the current and the magnetization. The data can be well fitted using this expression as shown in Figure 4c and (d). The obtained value of parameter $\alpha$ is $2.6 \times 10^{-21}$ m³Oe/W. The averaged $\beta$ of the device with $t_{Ta}=1.5$ nm is +/-9.2 Oe/($10^7$Acm⁻²), where +(-) indicates the device was pre-magnetized by

$H_x$= -(+)200 Oe. $H_{shift}$ was defined as $(H_{sw-}+H_{sw+})/2$, which linearly varies with a slope that depends of the direction of the pre-magnetizing field. $H_{shift}$ characterizes the effective magnetic field normal to the film plane, which has its magnitude linearly changed with the applied current and direction determined by the magnetization orientation of the top Co layer.

## 3. Conclusion

In summary, without external magnetic field, we have realized spin orbit torque induced magnetization switching utilizing interlayer exchange coupling in Ta(3)/Pt(3)/Co(1.4)/Ta(1.5)/Co(6)/Pt(2) devices (thickness in nanometers). The two ferromagnetic Co layers were found to be antiferromagnetic exchange coupled. The antiferromagnetic exchange coupling field was found to be 630 ±5 Oe for $t_{Ta}$ = 1.5nm, which decreases with increasing the thickness of the Ta layer. The current-induced switching behavior can be tuned by changing the pre-magnetization the in-plane Co layer. The magnitude of the current-induced perpendicular effective magnetic field from spin-orbit torque is about 9.2 Oe/($10^7$Acm$^{-2}$). Due to the large spin Hall angle of the Ta layer, compared to previously studied devices with Ru as spacer, the critical current density ($9\times10^6$A/cm$^2$) of the current-induced magnetization switching was reduced roughly by a factor of 3.

## 4. Experimental Section:

*Sample Preparation:* The stack structures of Ta(3)/Pt(3)/Co(1.4)/Ta($t_{Ta}$)/Co(6)/Pt(1.5) (thickness in nanometers) were deposited on Si/SiO2 substrates by magnetron sputtering, where $t_{Ta}$ is 1.5, 2, 3 and 5 nm, respectively. Sample Ta(3)/Pt(3)/Co(1.4)/Ta(1)/Pt(2) without top ferromagnetic layer is also studied as a reference. The deposition rates for Ta, Pt, Co films were controlled to be ≈0.018 nm s$^{-1}$, 0.027 nm s$^{-1}$ and 0.014 nm s$^{-1}$, respectively. The base pressure is less than 10$^{-8}$ Torr before deposition. The pressure of the sputtering chamber is 0.8 mTorr during deposition. Then the multilayers were patterned into Hall bar devices with channel width of 10 μm.

*Measurements:* The anomalous Hall effect measurements were carried out at room temperature with Keithley 2602B as the sourcemeter and Keithley 2182 as the nano-voltage meter. The Kerr measurement was taken using a NanoMoke3 magneto optical magnetometer.

**Acknowledgements:**

This work was supported by National Key R&D Program of China No.2017YFA0303400 and 2017YFB0405700. This work was supported also by the NSFC Grant No. 11474272, and 61774144. The Project was sponsored by Chinese Academy of Sciences, grant No. QYZDY-SSW-JSC020, XDPB0603, XDPB0802 and K. C. Wong Education Foundation as well.